# Low-temperature ageing of zirconia-toughened alumina ceramics and its implication in biomedical implants

Sylvain Deville[a], Jérôme Chevalier[a], Gilbert Fantozzi[a], José F Bartolomé[b], Joaquín Requena[b], José S Moya[b], Ramón Torrecillas[c], Luis Antonio Díaz[c]

[a] Institut National des Sciences Appliquées (INSA), GEMPPM, UMR 5510, 20 av. Albert Einstein, 69621 Villeurbanne, France
[b] Instituto de Ciencia de Materiales de Madrid (ICMM), CSIC, Cantoblanco 28049, Madrid, Spain
[c] Instituto Nacional del Carbón CSIC, C/ Francisco Pintado Fe, 26 La Corredoira, 33011 Oviedo, Spain

**Abstract**

Changes in crystalline phases resulting from low-temperature ageing of different yttria doped and non-doped zirconia-toughened alumina composites and nanocomposites were investigated under controlled humidity and temperature conditions in autoclave. A classical powder mixing processing route and a new modified colloidal processing route were used to process the composites. Different compositions ranging from 2.5 wt.% zirconia in a matrix of alumina to pure zirconia (3Y–TZP) were studied. It was observed that $Al_2O_3$+yttria stabilised $ZrO_2$ composites exhibited significant ageing. However, ageing was much slower than traditionally observed for Y–TZP ceramics, due to the presence of the alumina matrix. Ageing was clearly limited for zirconia content beyond 25 wt.%. On the other side of the spectrum, $Al_2O_3$+2.5 wt.% $ZrO_2$ initially presented a monoclinic fraction but did not show any ageing degradation. These composites seem to represent the best choice between slow crack growth and ageing resistance.

## 1. Introduction

Among the candidates for hip joint heads in orthopaedic applications, ceramic materials are the best choices so far. Their very high modulus, high compressive strength and the good compromise they represent between fracture resistance and wear behaviour have quickly drawn attention of the biomaterials world. Since metal-based solutions for hip joints prostheses have been questioned due to wear and corrosion problems, the development of new advanced ceramics is of growing interest. The mechanical properties of alumina and its in vivo stability have made it a good alternative in total hip replacement (THR) for some time,[1,2] but the non-negligible fracture rate is still limiting the reliability of implants. Since Garvie et al.[3] reported first the phase transformation reinforcement possibilities of zirconia, an extensive work has been performed on the development of zirconia-based ceramics. Zirconia ceramics present a unique combination of mechanical properties: high wear resistance, low coefficient of friction and higher crack resistance than alumina and very good biocompatibility, making them very attractive as far as orthopaedic applications are concerned.[4-6] This increase in crack resistance is accounted for by its ability to undergo phase transformation[7-12] from metastable tetragonal to monoclinic phase at room temperature, accompanied by a 4 vol.% increase but also microcracking. The resulting compressive stresses around the transformed zone and the energy spent for the transformation and microcracking provide a high fracture resistance. However, when unaccomodated, the volume increase can result in catastrophic failure[13] of the ceramic component, making them unreliable6 for orthopaedic applications. This phase transformation also results in some degradation of materials properties such as strength.[14,15] This degradation phenomenon is known as ageing.

The main features of ageing are summarised[16] as follows:
- Transformation proceeds most rapidly at 200–300 °C and is time dependent.
- The degradation is caused by the t–m transformation, accompanied by microcracking.
- Water or water vapour enhance the transformation.
- Transformation proceeds from the surface to the bulk of zirconia materials.
- Higher stabilising content or finer grain size increase the resistance to transformation.

A great amount of work has been dedicated to understand the underlying mechanisms of the ageing phenomenon's associated with zirconia, but views are still diverging.[16–20]

Ageing is clearly a limiting factor of zirconia use in orthopaedic applications. However, several possibilities of improving zirconia reliability are currently under investigations, e.g. silica as an additive.[21,22] The recent development of alumina-zirconia composites and nanocomposites appears as a good alternative: the mechanical properties are improved compared to both alumina and zirconia,[23-26] e.g. fracture toughness and slow crack growth propagation threshold,[27] and ageing phenomenon could be minimised if not completely avoided.[28] The influence of small amounts of alumina particles on zirconia ageing have been studied by several authors.[29–32] But it seems that little attention has been paid so far to the ageing mechanisms of zirconia-toughened alumina (ZTA) composites, where zirconia is the minor phase, though reports[14,28,32] are found on the degradation of mechanical properties under hydrothermal conditions. Very little interest has been paid to the zirconia content limit in ZTA composites beyond which ageing is completely avoided, provided this limit exists. The closest relevant value could be found in the work of Tsukuma and Shimada,[28] where the phase transformation is somewhat restrained to 15 wt.% by the addition of 40 wt.% of $Al_2O_3$ in 2Y–PSZ. However, there are no report yet where ageing was completely avoided.

This paper presents the study of ageing phenomenon in alumina–zirconia composites of several compositions, ranging from 2.5 wt.% zirconia in a matrix of alumina to pure zirconia. The influence of the yttria content and microstructure on the ageing mechanisms is also discussed.

## 2. Materials processing

Materials (Table 1) were processed either by conventional powder-mixing route or by a modified colloidal processing route involving zirconium–alkoxides.

| Designation | Compositions | Zirconia vol.% | Processing route | Sintering temperature |
| --- | --- | --- | --- | --- |
| A10TZY | $Al_2O_3$–10 wt.% ($ZrO_2$–$3Y_2O_3$) | 6.7 | Mixing powder | 1600 °C |
| A15TZY | $Al_2O_3$–15 wt.% ($ZrO_2$–$3Y_2O_3$) | 10 | Mixing powder | 1600 °C |
| A20TZY | $Al_2O_3$–20 wt.% ($ZrO_2$–$3Y_2O_3$) | 13.3 | Mixing powder | 1600 °C |
| A25TZY | $Al_2O_3$–25 wt.% ($ZrO_2$–$3Y_2O_3$) | 16.7 | Mixing powder | 1600 °C |
| A30TZY | $Al_2O_3$–30 wt.% ($ZrO_2$–$3Y_2O_3$) | 20 | Mixing powder | 1600 °C |
| A60TZY | $Al_2O_3$–60 wt.% ($ZrO_2$–$3Y_2O_3$) | 40 | Mixing powder | 1600 °C |
| A2,5Z-co | $Al_2O_3$–2,5wt.% $ZrO_2$ | 1.7 | Colloidal | 1600 °C |
| A10Z-co | $Al_2O_3$–10 wt.% $ZrO_2$ | 6.7 | Colloidal | 1600 °C |
| A10Z | $Al_2O_3$–10 wt.% $ZrO_2$ | 6.7 | Mixing powder | 1600 °C |
| A15Z | $Al_2O_3$–15 wt.% $ZrO_2$ | 10 | Mixing powder | 1600 °C |
| 3Y–TZP | $ZrO_2$–$3Y_2O_3$ | 100 | Mixing powder | 1500 °C |

Table 1. Designation and composition of the samples

Concerning the conventional processing route, a high purity alumina powder α-$Al_2O_3$ >99.9 wt.% (Condea HPA 0.5, Ceralox division, Arizona, USA) was mixed with different amounts (10–60 wt.%) of either yttria-stabilised zirconia powder (3Y–TZP, Tosoh TZ-3YS, Tosoh corporation, Tokyo, Japan) or unstabilised zirconia powder (Tosoh TZ-O, 10–15 wt.%).

$Al_2O_3$ powders with 2.5 wt.% and 10 wt.% of $ZrO_2$ was also processed through a modified colloidal processing route developed by Schehl et al.[33] Stable suspension of the above mentioned alumina powders in absolute ethanol (99.97%) were doped by dropwise addition of a diluted zirconium alkoxide solution.[34,35] After magnetic stirring at 60 °C and drying at 120 °C, the powders were thermally treated at 850 °C for 2 h and subsequently attrition milled, as a suspension in alcohol, with 3 mm alumina balls for 1 h. The powders were dried and sieved to less than 45 μm.

Different stable aqueous suspensions of 70 and 65 wt.% solids content using 0.5 and 1 wt.% addition of an alkali-free organic polyelectrolite (Dolapix C64), with the conventional and modified powders mixtures respectively, were obtained. The suspensions were highly dispersed, exhibiting pseudo-plastic flow behaviour and low viscosity ($\tau \leqslant 10$ mPas at shear rate of 500 s−1). Before carrying out the measurements, bubbles were removed by slow speed agitation. Ball milling was then performed using high purity alumina balls in an alumina jar for 24 h. Plates (100×100×5 mm) were cast from each suspension in plaster of Paris mould and dried in air at room temperature for 48 h. Samples with alumina content were then sintered at 1600 °C for 2 h in air. Pure zirconia samples were also elaborated from yttria-stabilised zirconia powder (Tosoh TZ-3YS). In this particular case, powders suspensions were slip-casted and then sintered at 1500 °C for 2 h. It is worth noting that all materials exhibit a density higher than 97%. Every composition studied here has the potential for being used as bioceramic for orthopaedic applications.

Sintered plates were then machined to small bars of 6 mm×4 mm×50 mm. The side of each bar on which XRD-analysis was to be performed were then mirror-polished using diamond slurries and pastes (diamond paste and slurries, PRESI, France).

## 3. Experimental procedure

### 3.1. Ageing experiments

Since classical surgery sterilisation has been extensively performed at a temperature of 134 °C in steam, most of the ageing experiments are usually carried out on biomedical grade zirconia materials at this temperature. Great concern is attached to the effect of sterilization on prostheses lifetime and zirconia degradation phenomenon rising from this sterilization,[36,37] whose effects have to be added to the in vivo degradation at body temperature.

Here, ageing experiments were carried out using a steam autoclave at 140 °C, with two bars pressure. In the present work, this temperature was chosen to reduce experimentation times, since ZTA composites exhibit significant lower ageing rate than 3Y–TZP.

Polished samples were put in steam autoclaves (Fisher Bioblock Scientific, France) and left in steam atmosphere for different times (up to 115 h). Samples were located on a grid in the autoclave so that they were not soaked in water during ageing, but only subjected to steam atmosphere. Before increasing the temperature up to 140 °C, enclosure was left open to evacuate the air atmosphere initially present in the steam autoclave, thus ensuring a 100% humidity atmosphere during ageing.

## 3.2. X-ray diffraction phase analysis

X-ray diffraction data were obtained with a diffractometer using Ni-filtered Cu-Kα radiation. The tetragonal/monoclinic zirconia ratio was determined using the integrated intensity (measuring the area under the diffractometer peaks) of the tetragonal (101) and two monoclinic (111) and (−111) peaks as described by Garvie and Nicholson,[38] and then revised by Toroya et al.[39,40] For the purpose of comparison, the obtained integrated intensities were individually normalised to the (101) tetragonal integrated intensity for each composition. Diffractometer scans were obtained from 27° to 33°, at a scan speed of 0.2°/min and a step size of 0.02°.

## 3.3. SEM imaging

SEM micrographs were obtained using a Philips XL20 with an accelerating tension of 25 kV. Polished samples were preliminarily thermally etched in air atmosphere at 1450 °C for 30 min, with heating and cooling rates of 400 °C/h. Samples were then gold-coated before SEM analysis. Aged samples were not thermally etched prior SEM analysis.

## 4. Results

## 4.1. Microstructures

The microstructures obtained from powder mixing route are quite similar. They are formed (Fig. 1) of a matrix of alumina grains (for compositions ranging from 10 to 30 wt.% zirconia content) and dispersed zirconia grains. For all compositions (yttria-stabilised and unstabilised), a significant amount (roughly 30 vol.%) of aggregates (up to 10 μm) of zirconia grains is present. They are originating from the starting powder.

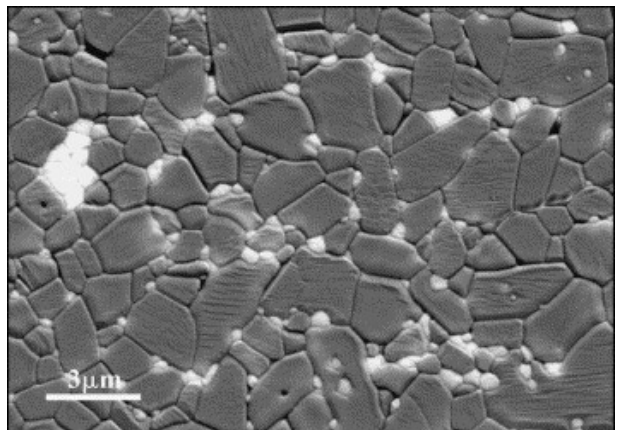

Fig. 1. A10TZY. Zirconia is the white phase. The presence of quite large aggregates should be noted.

As far as the samples fabricated by the colloidal processing route are concerned (2.5 and 10 wt.% zirconia), a very fine dispersion of zirconia grains within the alumina grains matrix can be observed (Fig. 2). The zirconia grain size is around 200 nm, several times finer than the alumina grain size. Those zirconia particles are found to be both intergranular, situated at either grain boundaries or triple points, and intragranular, as indeed some zirconia particles are found within alumina grains.

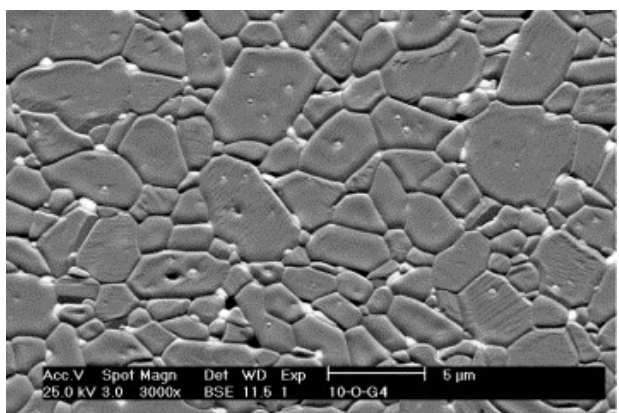

Fig. 2. A2.5Z-co, colloidal route. The very fine microstructure shows no aggregate at all.

### 4.2. XRD phase analysis

### 4.2.1. ZTA composites and nanocomposites

For all the compositions of yttria-stabilised composites, the starting surfaces after polishing are free of monoclinic phase. After 2 h of ageing at 140 °C, all these composites show an increase of monoclinic phase fraction up to 10%, as shown on Fig. 3. Two different behaviours could then be observed. For low zirconia content (e.g. <20 wt.%), almost no more evolution of the monoclinic phase fraction is found. Variations are within the measurement error range. The maximum monoclinic phase fraction reached after 115 h at 140 °C is about 15%. On the other hand, when the zirconia content is high enough (e.g. >25 wt.%), the monoclinic fraction is still increasing, at quite a steady rate. This second stage occurs however at a lower rate than for the first 2 h. It should also be noted that the higher the zirconia content, the higher the ageing rate.

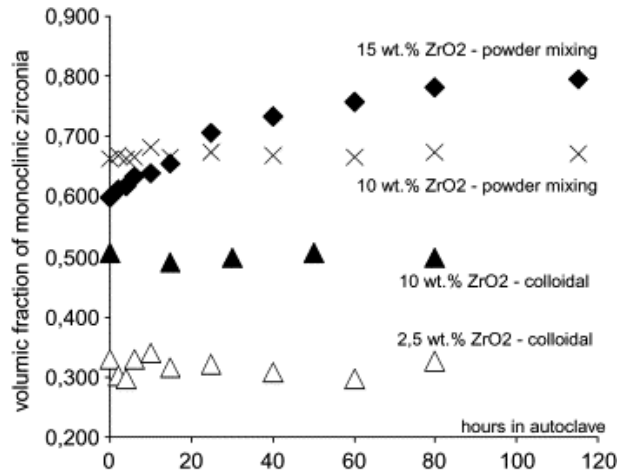
Fig. 3. Monoclinic phase fraction evolution for the unstabilized materials.

The monoclinic phase volume fraction evolution for the unstabilised materials is shown on Fig. 4. The starting monoclinic phase fraction after polishing are, respectively, of 66 and 60% for the A10Z and A15Z. However, during ageing, two different behaviours could be observed. For the 10 wt.% zirconia, no phase fraction evolution is observed as a function of ageing time. Phase fraction variations are within the measurement error range. The monoclinic phase fraction is stable around 66%, value reached immediately after polishing, before any ageing was performed yet. As far as the A15Z is concerned, the starting monoclinic phase fraction is around 60% and increases at a quite steady rate with ageing time, reaching 80% after 115 h of ageing at 140 °C.

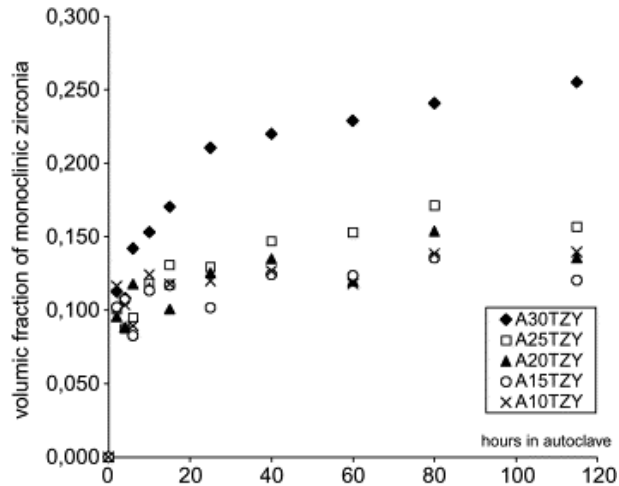
Fig. 4. Monoclinic phase fraction evolution for the yttria stabilized materials: influence of zirconia content.

### 4.2.2. 3Y–TZP

The ageing behaviour of 3Y–TZP has been extensively studied by many authors.[17–20,41] It should be noted that the ageing rate obtained here (Fig. 3) is much faster than for all the ZTA composites. The maximum monoclinic phase fraction is increasing very fast up to the maximum value (~80 vol.%), that is reached after only 20 h at 140 °C. This already demonstrates that ZTA ceramics, whatever their compositions, exhibit much better ageing resistance than monolithic 3Y–TZP.

## 5. Discussion

### 5.1. Unstabilised materials

As shown on Fig. 4, two different behaviours can be observed for the unstabilised materials: either ageing is occurring at quite a steady rate, or no ageing at all is observed, depending on the zirconia content. This behaviour might be related to the microstructure in a fairly simple manner. Fracture toughness behaviour of alumina-unstabilised zirconia composites has been the object of an extensive work in the past, and it has been shown[42] there is an increase of fracture toughness while increasing the zirconia content up to about 10 vol.%. Indeed, increasing the number of zirconia particles increases the number of microcracks created during cooling after sintering and when cracks are propagating within the material. The toughness is therefore increased, by microcracking toughening. However, beyond a certain level, those microcracks tend to percolate altogether, joining up between the particles, and the toughness is abruptly decreasing. This microcracks percolation may also account here for the ageing behaviour. Microcracks present along the grain boundaries of the alumina matrix could act as preferential paths for the water diffusion inside the bulk of the material. The tensile stress accompanying the t–m transformation upon cooling and the presence of water, cause microcracks to grow at subcritical stress level. These microcracks change the transformation conditions of the adjacent tetragonal grains, such as strain energy and strain force between the grains, etc. They consequently contribute to accelerate the transformation of tetragonal phase to monoclinic during ageing, generating additional microcracks and macrocracks on the transformed surface owing to volume expansion upon transformation, and leading to the severe degradation of mechanical properties. When the zirconia content is above the mentioned maximum, the microcracks percolation system will provide the ceramic with preferential paths for water propagation within the material, increasing the ageing degradation.

However, by keeping the zirconia content beyond this maximum, unstabilised zirconia particles will transform to monoclinic during cooling, without percolation of microcracks. It should be emphasised that in absence of yttria, no ageing degradation was observed. The monoclinic phase fraction is kept constant at its starting value reached after processing.

## 5.2. Yttria-stabilised materials

The two stages observed for the yttria stabilised materials could be related to two distinct phenomenon. The first stage, occurring during the first 2 h of ageing at 140 °C is characterised by an sudden increase of monoclinic fraction from 0 to about 10 wt.% for all compositions. Since this increase does not seem to depend on the zirconia content, it should be related to some starting powder characteristics. The SEM observations (Fig. 1) show the presence of aggregates of zirconia grains within the alumina matrix. Although these aggregates tend to be scarce, there is a large number of grains involved in each one of them. A rough estimation of the volume fraction of aggregates leads to values around 30 vol.% of the zirconia phase. Zirconia grains within those aggregates, since they are surrounded by other zirconia grains with the same modulus, might behave roughly in the same manner than in a pure zirconia. Therefore most of them could transform during the first stage of ageing, leading to the observed increase. The presence of an higher strain energy barrier induced by the alumina grains would then prevent the transformation of the neighbouring zirconia grains. This could explain why only a third of the agglomerates is transforming. A deeper investigation will be performed, using atomic force microscopy techniques to validate this hypothesis.

The second stage of ageing is characterised by the continuation or not of the ageing degradation, depending on the zirconia content, as opposed to the first stage. This behaviour could again be related to the microstructure when considering the ageing propagation mechanism. It should be reminded that the t–m phase transformation is accompanied by the creation of microcracks. The induced microcracks surrounding the transformed grain will ease the propagation of water within the material, and all the surrounding grains will therefore tend to transform as well. Once initiated, the degradation will thus proceed from one grain to another. Provided the zirconia grains repartition is homogeneous and that the zirconia content is low enough, all the zirconia grains will be separated from each other. If one grain transforms, ageing will not be able to proceed, since no contact with another zirconia grain is found. An insignificant degradation is then observed with ageing time (e.g. A10TZY), with a monoclinic phase fraction reaching values only around 12.5% after 115 h.

On the other hand, when the zirconia content is high enough (e.g. >25 wt.%), percolation of zirconia grains is reached (~22 wt.%).43 The existence of paths of zirconia grains will ease the ageing degradation propagation. Since the zirconia grains are initially stabilised by yttria, there is no microcrack present within the material after processing, as opposed to the unstabilised materials. If zirconia grains are not in contact with each other, microcracks induced by the transformation of one grain will not be sufficient for the water to diffuse within the material to the neighbouring zirconia grains. When percolation paths are present, the microcracks created by phase transformation will allow diffusion of water to the next surrounding grains so that ageing is able to propagate.

Nucleation and growth processes of the transformed zones are occurring at the same time, leading to a higher rate of degradation. However, the surrounding alumina grains still have a higher bulk modulus than the zirconia grains. This effect will moderate the phase transformation.[8,44] The lower the zirconia content, the slower the ageing rate. When increasing the zirconia content, more and more zirconia grains are found, and the transformation becomes easier. The observed ageing rates in ZTA are indeed much slower than in 3Y–TZP.

### 5.3. Interest of colloidal processing route

The influence of the microstructure homogeneity has been hypothesised here. However, using classical powder-mixing processing route to reach a very fine and homogeneous microstructure has been proved to be almost impossible target. Avoiding aggregates formation during the process is a difficult task, when considering all the processes developed so far. On the other hand, the SEM study of the microstructure of samples obtained by the colloidal processing route clearly show a very fine and also very homogeneous microstructure, with no evidence of the presence of aggregates (Fig. 2). With such a processing route, it should be possible to increase the zirconia content up to values that will lead to better mechanical properties, but avoiding ageing phenomena related to the presence of aggregates. The advantages of such a microstructure is double: by keeping a nanometer grain size, the important residual strains after cooling will improve the resistance to crack propagation[44] and thus toughness. By avoiding zirconia grains aggregates, it would be possible to increase the zirconia volume fraction so that the transformation toughening becomes effective in the material and therefore increases the toughness again. Compared to the unstabilised materials fabricated by the powder mixing route, the starting monoclinic phase fraction could be much lower, therefore keeping all the potential for transformation toughening. The A10Z and A15Z samples start with monoclinic volume fraction as high as 60 or 66 vol.%, which limits the potential for transformation toughening.

### 5.4. Surface degradation

As far as hip prostheses applications are concerned, surface roughness as low as a few nanometers is required, in order to limit the amount of debris during wear. Small fluctuations of monoclinic phase content will not have tremendous effects on the material properties. However, the surface degradation resulting from phase transformation is critical issue as far as the wear behaviour is concerned. This in vivo surface degradation could be anticipated from the SEM micrograph (Fig. 5) taken in our study after 115 h of ageing. Even without any thermal etching, some grains pull-outs are observed. The microcracks surrounding the transformed grains led to these pull-outs, and indeed significant number of them can be observed. Provided enough microcracks are created, it should be possible that even alumina grains would be subjected to pull-out (Fig. 6). The presence of these grains would be very damageable in terms of third body wear and of the debris generated in the surrounding tissues. Their micrometer-size has to be compared with the starting surface roughness of cups and heads, in the order of a few nanometers. Once this phenomenon is initiated, the size of the damaged zone is increasing and is able to reach sizes in the range of a few dozen of microns Fig. 7. The consequences of small ceramics debris on surrounding tissues and implants lifespan are not yet clearly understood, but they might be an important problem. As far as biomedical applications are concerned, it is therefore necessary to avoid completely any ageing phenomenon. Even a small increase in monoclinic phase fraction could be very damageable in terms of wear and biological tissues response.

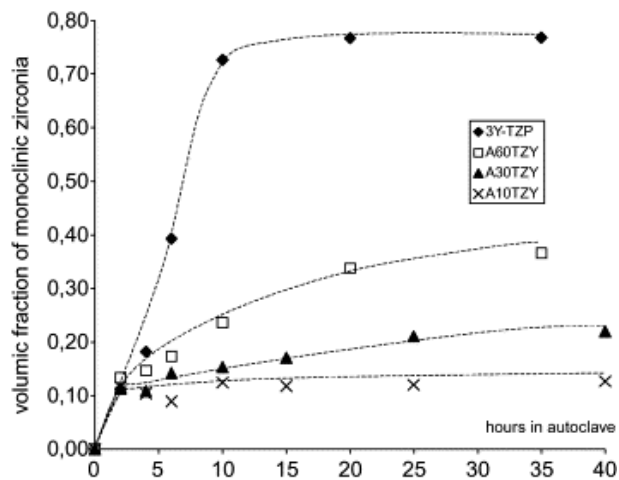

Fig. 5. Monoclinic phase fraction evolution of yttria-stabilised materials: comparison with 3Y–TZP.

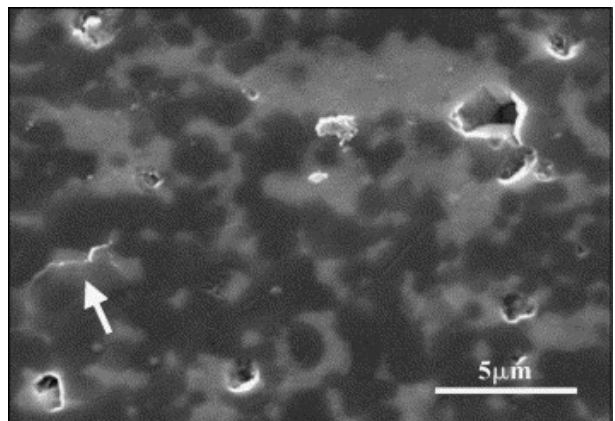

Fig. 6. Grain pull-outs and microcracking (arrow) in an A30TZY after 115 h of ageing at 140 °C. Zirconia is the white phase.

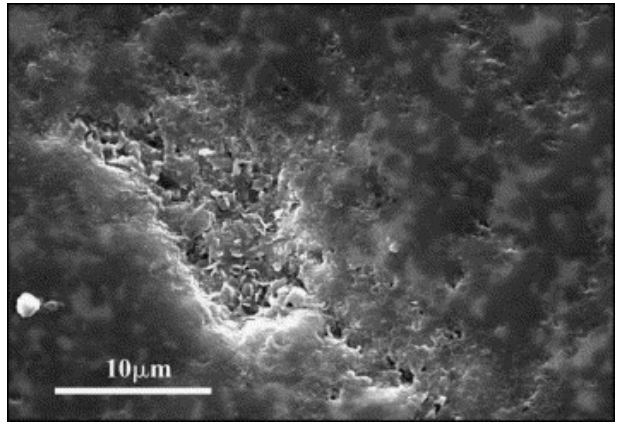

Fig. 7. SEM micrograph of surface degradation in a A30TZY after 115 h of ageing at 140 °C. Lots of microcracks and grains pull-outs can be observed.

## 6. Conclusions

1. As far as the unstabilised zirconia containing materials are concerned, ageing seems to be occurring by water diffusion within the cracks created during cooling, at the end of processing. If the zirconia content is kept low enough, those microcracks are not percolating, thus water is not able to diffuse and the material shows no evidence of ageing. As far as biomedical applications are concerned, zirconia content must be kept beyond 6.7 vol.% (i.e. 10 wt.%) in order to avoid ageing degradation.
2. Concerning yttria-stabilised zirconia containing materials, the monoclinic phase fraction starts from zero, but the presence of yttria allows ageing to proceed if the zirconia content is chosen beyond the percolation point. However, ageing is much slower than traditionally observed for Y–TZP ceramics. To avoid ageing degradation, it is therefore necessary to choose zirconia content below the percolation threshold, i.e. below 16 vol.% (i.e. 22 wt.%), and also probably to avoid aggregates formation within the material, which is very difficult by using a powder mixing processing routes. The transformation of the zirconia grains aggregates may cause some additional degradation, leading to grain pull-outs and surface degradation that would be very damageable in terms of wear behaviour and body reaction.

Looking for the best compromise between ageing sensibility and mechanical properties, the modified colloidal route seems to offer very promising results. It has been shown that it is possible to reach very fine and very homogeneous microstructures, without yttria and also without any aggregates. Since their mechanical properties are also very interesting,[23–27] those materials would therefore be the best choice, as far as biomedical applications are concerned. Any ageing sensibility should be absolutely avoided.

## Acknowledgements

The authors would like to acknowledge the EU for the financial support under the GROWTH2000, project reference GRD2-2000- 25039.